\documentclass[10pt,letterpaper,twocolumn]{article} 

\usepackage{ol2}
\usepackage[draft]{hyperref}
\usepackage{amsmath}
\usepackage{float}
\usepackage{graphicx}
\usepackage[nooneline]{subfigure}

\def\be{\begin{equation}}
\def\ee{\end{equation}}

\begin{document}

\twocolumn[ 

\title{A Precision Angle Sensor using an Optical Lever inside a Sagnac Interferometer}

\author{J. M. Hogan,$^{\dagger}$ J. Hammer,$^{\dagger}$ S.-w. Chiow, S. Dickerson, D. M. S. Johnson, T. Kovachy, A. Sugarbaker and M. A. Kasevich$^*$}

\address{Department of Physics, Stanford University, Stanford, California 94305, USA\\
$\dagger$Primary contributors.
$^*$Corresponding author: kasevich@stanford.edu}
\begin{abstract}We built an ultra low noise angle sensor by combining a folded optical lever and a Sagnac interferometer.  The instrument has a measured noise floor of  $1.3~\text{prad}/\sqrt{\text{Hz}}$ at $2.4~\text{kHz}$.  We achieve this record angle sensitivity using a proof-of-concept apparatus with a conservative $N=11$ bounces in the optical lever.  This technique could be extended to reach sub-$\text{picoradian}/\sqrt{\text{Hz}}$ sensitivities with an optimized design.
\end{abstract}


 ] 

Recently, a single-loop Sagnac interferometer has been shown to have exceptional performance as a precision angle sensor \cite{DixonPRL}.  In this Letter, we show how this performance can be substantially improved by incorporating an optical lever into the interferometer topology.  The resulting sensor has achieved unprecedented noise performance levels in our proof-of-concept work.  We anticipate that future sensors based on this method could achieve a million-fold improvement over the current state-of-the-art.  In addition to technological applications in, for example, precision beam steering systems, such sensors may achieve the sensitivities required to play a role in the detection of gravitational waves \cite{TOBA}.

The optical lever is a well-known precision angle deflection sensor.  Previously, an optical lever was used to demonstrate an angle noise floor of $10~\text{prad}/\sqrt{\text{Hz}}$ \cite{Lorrain}.  In a folded optical lever, a laser beam reflected $N$ times off a mirror will deflect by an angle $2N\delta\theta$ in response to an angular change $\delta\theta$ of the mirror, giving an $2N$-fold enhancement to the angular sensitivity of the measurement.  Alternatively, recent results have shown that a laser Sagnac interferometer can serve as a precision angle sensor \cite{DixonPRL}.  When aligned near the condition of complete destructive interference (the dark port condition), the Sagnac spatial interference pattern becomes increasingly sensitive to angular deflections.  Although the fundamental limit of the Sagnac configuration is the same as the optical lever deflection technique \cite{Putman}, the Sagnac dark port enhancement can lead to the suppression of certain classical noise sources \cite{StarlingSNR}.  Also, unlike a deflection measurement, the Sagnac geometry is intrinsically insensitive to fluctuations in the input angle.

\begin{figure}
\begin{center}
\includegraphics[width=2.5 in]{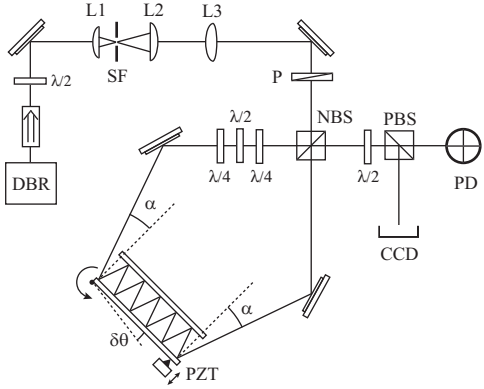}
\caption{ \label{Fig:Setup} Experimental setup consisting of an optical lever inside a Sagnac interferometer.  Deflections $\delta\theta$ of one of the optical lever mirrors are measured using a position sensitive split detector (PD). SF: pinhole spatial filter. PZT: piezoelectric transducer. NBS: non-polarizing beamsplitter. PBS: polarizing beamsplitter. P: polarizer. $\lambda/4$: quarter waveplate. $\lambda/2$: half waveplate. L1, L2, L3: lenses. DBR: $850~\text{nm}$ laser source.}
\end{center}
\end{figure}

Angular deflections of one of the mirrors inside a Sagnac interferometer cause the clockwise (CW) and counter-clockwise (CCW) interferometer beams to deflect in opposite directions at the output of the interferometer (see Fig. \ref{Fig:Setup}).  This results in a spatial interference pattern that translates in response to angular deflections by an amount that depends on the relative phase $\phi$ between the CW and CCW beams \cite{HowellPRA}.  Since small shifts of the interference pattern approximately translate the beam, a position sensitive detector at the output can measure the angular deflection of the mirror.  For a split detector measuring powers $p_1$ and $p_2$, the normalized first-order response to a beam deflection of angle $\delta\theta$ is
\be S=\frac{p_1-p_2}{p_1+p_2} = \sqrt{\frac{2}{\pi}}k\sigma \delta\theta \cot\!{(\phi/2)} \label{Eq:SplitDetectorSagnacResponse}\ee
where $k=2\pi/\lambda$ is the wavevector of the light, $\sigma$ is the $e^{-2}$ radial waist of the (assumed Gaussian) interferometer beam, and the beam waist is taken to be much smaller than the characteristic size of the detector.  Notice that Eq. (\ref{Eq:SplitDetectorSagnacResponse}) implies that as $\phi \rightarrow 0$, corresponding to complete destructive interference between the CW and CCW beams, the response $S$ diverges.  This enhancement near the dark port condition can yield substantial improvements in angular precision.  Such a protocol has recently been interpreted in the context of weak measurement \cite{DixonPRL}.

The optimum Sagnac phase $\phi$ depends on the character of the noise present in the instrument \cite{StarlingSNR}.  This can be seen by considering the signal to noise ratio ($\text{SNR}$) for several classes of typical noise.  For a split detector, the difference signal is $\Delta p=(p_1-p_2)\propto k \sigma^3 \delta\theta \sin\!{(\phi)}$ whereas the total power incident on the detector is $p_t=(p_1+p_2)\propto \sigma^2 \sin^2\!{(\phi/2)}$ so that Eq. (\ref{Eq:SplitDetectorSagnacResponse}) follows from $\Delta p/p_t$.  Assuming the noise scales as $\delta\left(\Delta p\right) \propto (p_t)^n$ for some power $n$ we have $\text{SNR}=\tfrac{\Delta p}{\delta\left(\Delta p\right)}\propto\sin\!{(\phi)} \sin\!^{-2n}{(\phi/2)}$.  For example, certain technical noise sources (stray light, electronics noise, etc.) are modeled by $n=0$ and result in a peak $\text{SNR}$ at $\phi=\pi/2$.  In the case of photon shot noise ($n=\tfrac{1}{2}$), the SNR approaches its optimum value as $\phi\rightarrow 0$.  For any noise source with $n>\tfrac{1}{2}$, including classical intensity noise ($n=1$), the SNR diverges as $\phi\rightarrow 0$.  Optimizing the value of $\phi$ in the given noise environment reduces the effective angle noise floor, moving the sensor closer to the shot-noise limit.

The experimental setup is shown in Fig. \ref{Fig:Setup}.  The Sagnac interferometer is formed using a non-polarizing beamsplitter cube (NBS) that directs incident light into the CW and CCW arms of the interferometer.  The light for the experiment is derived from an $850~\text{nm}$ DBR laser that we prepare with a pinhole spatial filter (SF) and a polarizer (P) to produce a Gaussian beam with horizontal polarization.  The beam size and divergence are adjusted by a set of lenses (L1, L2, L3) with focal lengths $f_1=75~\text{mm}$, $f_2=200~\text{mm}$ and $f_3=300~\text{mm}$, respectively.  The beam waist at the lens L3 is $\sigma_3 = 800~\mu\text{m}$.  Translating L3 adjusts the beam collimation, allowing the waist $\sigma$ at the detector to be varied from $600~\mu\text{m}$ to $2000~\mu\text{m}$.  The Sagnac interference pattern is monitored using a split photodetector (PD).  For diagnostic purposes, a small amount of light is directed onto a CCD camera to ensure a good mode overlap of the CW and CCW beams.

In place of one of the mirrors of the Sagnac interferometer, two nearly parallel mirrors are aligned to form an optical lever.  A piezoelectric transducer (PZT) mounted to one of the mirrors allows its angle to be adjusted by a controlled amount.  Typically we achieve up to $N=11$ bounces on the actuated mirror without substantial clipping of the beam at the mirror edges.  The input angle to the lever $\alpha\approx 12.5^\circ$ is chosen to maximize the number of bounces given the beam waist, the $10~\text{cm}$ mirror length, and the $d \approx 2~\text{cm}$ mirror separation.

\begin{figure}
\begin{center}
\includegraphics[width=3.2 in]{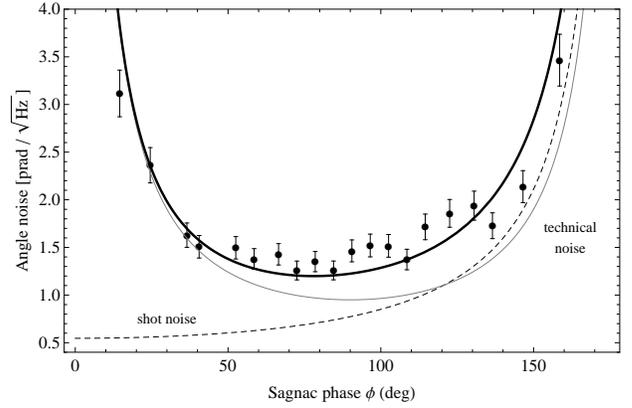}
\caption{\label{Fig:SensitivityVsPhi} Angle noise optimization. The fit (solid, dark) is to a noise model that includes shot noise (dashed) and technical noise (solid, light).}
\end{center}
\end{figure}

We characterize the Sagnac sensor scale factor $C_\text{sagnac}(\phi) = \delta\theta / S_\text{sagnac}(\phi)$ by applying a $2.4~\text{kHz}$ angle drive signal of known amplitude to the PZT mirror (see calibration procedure described below).  We then extract the signal noise floor from the RMS bandpower as measured by the split detector in a $150~\text{Hz}$ frequency band around the $2.4~\text{kHz}$ calibration signal.  The results adjusted by $C_\text{sagnac}(\phi)$ are shown in Fig. \ref{Fig:SensitivityVsPhi} as a function of the Sagnac phase $\phi$.  We obtain an optimum angular resolution of $(1.3\pm0.1)~\text{prad}/\sqrt{\text{Hz}}$ at $\phi= 84.6~\text{deg}$.   The fact that the $\text{SNR}$ is maximized near $90~\text{degrees}$ suggests that the measurement is dominated by technical noise.  The fit (solid) in Fig. \ref{Fig:SensitivityVsPhi} is to a hybrid noise model that includes technical noise ($n=0$) and shot noise ($n=\tfrac{1}{2}$), with the noise amplitudes as free parameters.

\begin{figure}
\begin{center}
\includegraphics[width=3.0 in]{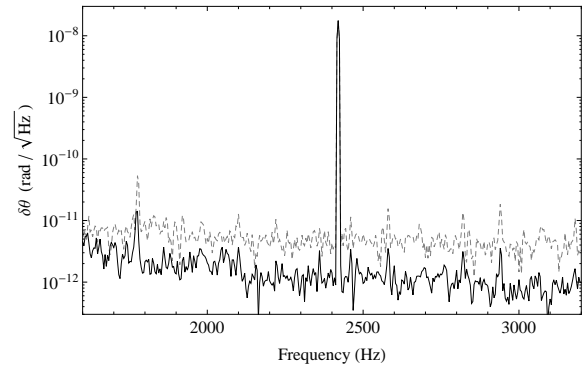}
\caption{ \label{Fig:ResolutionFFT} Angle amplitude spectral density.  The black (solid) curve is the Sagnac measurement for $\phi=84.6~\text{deg}$.  The gray (dashed) curve shows the deflection calibration trace with no Sagnac enhancement. The resolution bandwidth is $4~\text{Hz}$.}
\end{center}
\end{figure}

Figure \ref{Fig:ResolutionFFT} shows the angle noise amplitude spectral density of the sensor at the optimum resolution $\phi= 84.6~\text{deg}$ (black, solid) measured with an Agilent 35670A spectrum analyzer.  For comparison, the noise spectrum for a deflection measurement with the NBS removed is also shown (gray, dashed).  The spike at $2.42~\text{kHz}$ is the drive signal that we apply to the PZT for calibration.

In order to calibrate the PZT-actuated mirror, we perform a pure deflection measurement with the NBS element removed so that there is no interference.  For our optical lever geometry, an angular tilt $\delta\theta$ of the PZT mirror results in a beam deflection at the detector of $2 N \delta\theta L_\text{eff}$.  Here $L_\text{eff} \equiv L + \frac{d}{\cos^2{\alpha}}(N - 1)$ is the effective optical path length from the entrance of the lever to the detector which includes the deflection that occurs inside the lever itself, and $L\approx 40~\text{cm}$ is the distance from the output of the lever to the detector.  To characterize the response of the split detector to beam displacements, we translate the detector using a micrometer while keeping the beam fixed.  Within a range of $\sim 800~\mu\text{m}$ about the center, the normalized split detector signal $S_\text{defl}$ depends linearly on position, and we measure a slope of $R_\text{split}  = (1160 \pm 20)~\text{m}^{-1}$ for beam waist $\sigma=1.4~\text{mm}$.  Altogether, the calibrated deflection scale factor with the NBS removed is $C_\text{defl} \equiv \frac{\delta\theta}{S_\text{defl}} = (2 N L_\text{eff} R_\text{split})^{-1}= (64 \pm 2)~\mu\text{rad}$.

Next, still with the NBS removed, we drive the PZT mirror at $2.42~\text{kHz}$ with an amplitude of $22.5~\text{V}$.  The fractional split detector response to this drive is $S_\text{defl}=(5.54\pm 0.06)\times 10^{-4}$.  Using the scale factor $C_\text{defl}$, we infer an angular deflection of the mirror of $(35 \pm 1)~\text{nrad}$ at $2.42~\text{kHz}$.  Thus the PZT mirror provides a calibrated angular deflection per drive amplitude of $(1570 \pm 40)~\frac{\text{prad}}{\text{V}}$.

\begin{figure}
\begin{center}
\includegraphics[width=3.0 in]{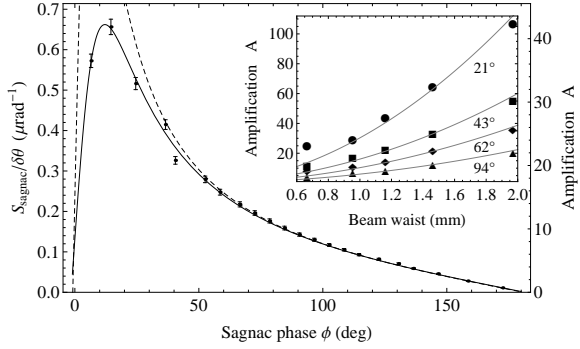}
\caption{\label{Fig:Enhancement} Sagnac signal response. The response initially increases as $\phi\rightarrow 0$ but is then suppressed due to intensity imbalance. For small $\phi$, theory using the measured intensity ratio $\eta=1.3$ (dashed curve) differs from the results of a single parameter fit ($\eta=1.61\pm 0.01$, solid curve).}
\end{center}
\end{figure}

Returning to the the Sagnac configuration, with the NBS reinserted, we measure the Sagnac response to the now calibrated angular deflection of the PZT mirror.  Fig. \ref{Fig:Enhancement} shows the measured fractional split detector response $S_\text{sagnac}$ for the same $22.5~\text{V}$ amplitude PZT drive signal at $2.42~\text{kHz}$ used in the deflection measurement.  The right axis of Fig. \ref{Fig:Enhancement} shows the amplification factor $A \equiv S_\text{sagnac}/S_\text{defl}$ obtained by comparison of the Sagnac signal to the deflection signal at $2.42~\text{kHz}$ as a function of the Sagnac phase $\phi$.

The Sagnac data in Fig. \ref{Fig:Enhancement} is fit using a more detailed theoretical model that allows for an intensity imbalance between the CW and CCW beams (\textit{e.g.}, from the imperfect NBS) as well as beam divergence\cite{HowellPRA}.  The predicted amplification factor is
\be A(\phi)=\frac{1}{L_\text{eff}}\frac{(\eta-1)L_\text{eff}+\sqrt{\eta}\, \gamma k \sigma^2 \sin{(\phi)}}{1 + \eta - 2 \sqrt{\eta} \cos{(\phi)}} \label{Eq:SagnacEnhancementA}\ee
where $\eta=I_\text{CW}/I_\text{CCW}$ is the intensity imbalance and the correction factor $\gamma=\frac{{l +  L_\text{eff}(\sigma_3/\sigma)}}{{l + L_\text{eff}}}$ accounts for beam divergence given a beam waist $\sigma_3$ at lens L3 a distance $l$ in front of the optical lever.  Consistent with our data, near $\phi=0$ the enhancement is suppressed for $\eta\neq 1$.

The inset of Fig. \ref{Fig:Enhancement} demonstrates the expected $\sigma^2$ dependence of the interferometric amplification as a function of the beam waist for several relative phases, including fits (gray) to the theoretical prediction of Eq. \ref{Eq:SagnacEnhancementA}.  The data in the inset was taken without an optical lever in order to simplify realignment and to avoid clipping the beam throughout the entire beam waist range.

Adjusting the phase $\phi$ between the CW and CCW beams requires a nonreciprocal element inside the interferometer.  Following Dixon \emph{et al.}\cite{DixonPRL}, we take advantage of the polarization degree of freedom of the light to implement the desired phase shift.  We insert a waveplate compensator into the Sagnac interferometer which consists of two quarter waveplates and one half waveplate (see Fig. \ref{Fig:Setup}).  The quarter waveplates are oriented with their fast axes at $45^\circ$ and $135^\circ$ with respect to the horizontal plane such that the horizontally polarized input light is rotated to vertical polarization after passing through the three waveplates in either direction.  The half waveplate imparts the desired differential phase $\phi=4\beta$ between the two directions, where $\beta$ is the angle between the fast axis of this waveplate and the horizontal plane.

The angle sensitivity achieved with this technique can be greatly improved.  The photon shot-noise limited angle sensitivity for a $10~\text{W}$ laser with a $1~\text{cm}$ beam waist and $N=10^3$ bounces in the optical lever would be $10^{-18}~\text{rad}/\sqrt{\text{Hz}}$.  Such a sensor could allow for scientifically interesting gravitational wave detection around $1~\text{Hz}$ with a strain sensitivity of $\sim\! 10^{-18}~/\sqrt{\text{Hz}}$ \cite{TOBA}.  This value of $N$ could be realized with large, low-loss mirrors and would potentially require arranging the beam spots in a two-dimensional grid on the lever mirrors.  The sensitivity would likely be limited by thermal fluctuations in the mirror substrate \cite{numata}.

\bibliographystyle{osajnl}

\end{document}